\documentclass[prl,preprint,showpacs]{revtex4}
\usepackage{graphicx}

\begin{document}
\title{Topological signature of deterministic chaos in short
  nonstationary signals from an optical parametric oscillator}

\date{July 30, 2003, revised January 6, 2004, to appear in Physical
  Review Letters}

\author{Axelle Amon and Marc Lefranc} \affiliation{Laboratoire de Physique des
  Lasers, Atomes, Mol\'ecules, UMR CNRS 8523, Centre d'\'Etudes et de
  Recherches Lasers et Applications, Universit\'e des Sciences et
  Technologies de Lille, F-59655 Villeneuve d'Ascq, France}

\begin{abstract}
  
  Although deterministic chaos has been predicted to occur in the
  triply resonant optical parametric oscillator (TROPO) fifteen years
  ago, experimental evidence of chaotic behavior in this system has
  been lacking so far, in marked contrast with most nonlinear systems,
  where chaos has been actively tracked and found. This situation is
  probably linked to the high sensitivity of the TROPO to
  perturbations, which adversely affects stationary operation at high
  power.  We report the experimental observation in this system of a
  burst of irregular behavior of duration 80 $\mu$s. Although the
  system is highly nonstationary over this time interval, a
  topological analysis allows us to extract a clearcut signature of
  deterministic chaos from a time series segment of only 9 base cycles
  (3 $\mu$s).  This result suggests that nonstationarity is not
  necessarily an obstacle to the characterization of chaos.

\end{abstract}
\maketitle

It has become common knowledge that many nonlinear systems obeying
deterministic equations of motion can display seemingly erratic
behavior. In the last decades, deterministic chaos has been the
subject of intensive experimental investigation and has been observed
in a large variety of experimental systems (see, e.g.,
\cite{cvitanovic84:_univer_chaos}). However, characterizing
chaos is significantly more demanding than characterizing periodic
behavior, for which measuring simple quantities such as amplitude or
frequency over a short time interval is sufficient. Indeed, most
quantitative measures of chaos (e.g., fractal dimensions or Lyapunov
exponents) rely on constructing an approximation of the natural
measure on a strange attractor, which requires observing the system
for at least a few hundreds of cycles at fixed control
parameters~\cite{Ott93,Abarbanel96}.

Thus, it is extremely difficult to assess deterministic chaos in a
system that
experiences parameter drifts on a time
scale comparable to the mean dynamical period, so that stationarity
cannot be assumed. In particular, this is the case when studying a
subsystem that cannot be considered as isolated from its environment,
a situation that frequently occurs in biological systems. Yet, it is
often desirable to understand the behavior of a small part of a
complex system before unraveling its global dynamics. A natural
question then is: can we infer the existence of an underlying chaotic
dynamics from a very short, nonstationary, time series? In this
Letter, we present a case in which this question can be answered
positively: by applying topological
tools~\cite{Gilmore02,Gilmore98,Boyland94:_topol} to a burst of irregular
behavior recorded in a triply resonant optical parametric oscillator
(TROPO) subject to thermal effects, we extract a clearcut signature of
deterministic chaos from an extremely short time series segment of
only 9 base cycles.

Optical parametric oscillators are sources of coherent light based on
parametric down-conversion of pump photons into pairs of subharmonic
photons in a nonlinear optical crystal. This effect is
enhanced by enclosing the crystal inside a cavity so as to build an
oscillator. When the cavity is resonant for the three waves, the
threshold for infrared generation can be as low as a few mW. As
lasers, TROPOs are based on a nonlinear interaction and are naturally
susceptible to instabilities and chaos. Accordingly, chaotic behavior
was identified in a simple TROPO model fifteen years
ago~\cite{Lugiato88}. Surprisingly, this theoretical prediction has so
far not been confirmed experimentally.

Instabilities and chaos are expected to occur in the TROPO at high
power, where optical nonlinearities are emphasized. However, high
energy densities in the crystal induce other effects, in
particular thermal effects. It was recently shown that the TROPO can
be subject to thermo-optical instabilities where the cavity length is
no longer a fixed parameter but behaves as a slow variable coupled to
the optical variables~\cite{Suret00}. This gives rise to relaxation
oscillations~\cite{Suret00,Suret01a}, as well as to a variety of
bursting regimes~\cite{amon03:_burst} when these slow oscillations
combine with fast oscillations resulting from the interaction of
transverse modes~\cite{Suret01b}. The coexistence of two different
time scales then makes it difficult to characterize the dynamics,
especially in the case of irregular regimes.

\begin{figure}[htbp]
  \centering
  \includegraphics[width=3.8in]{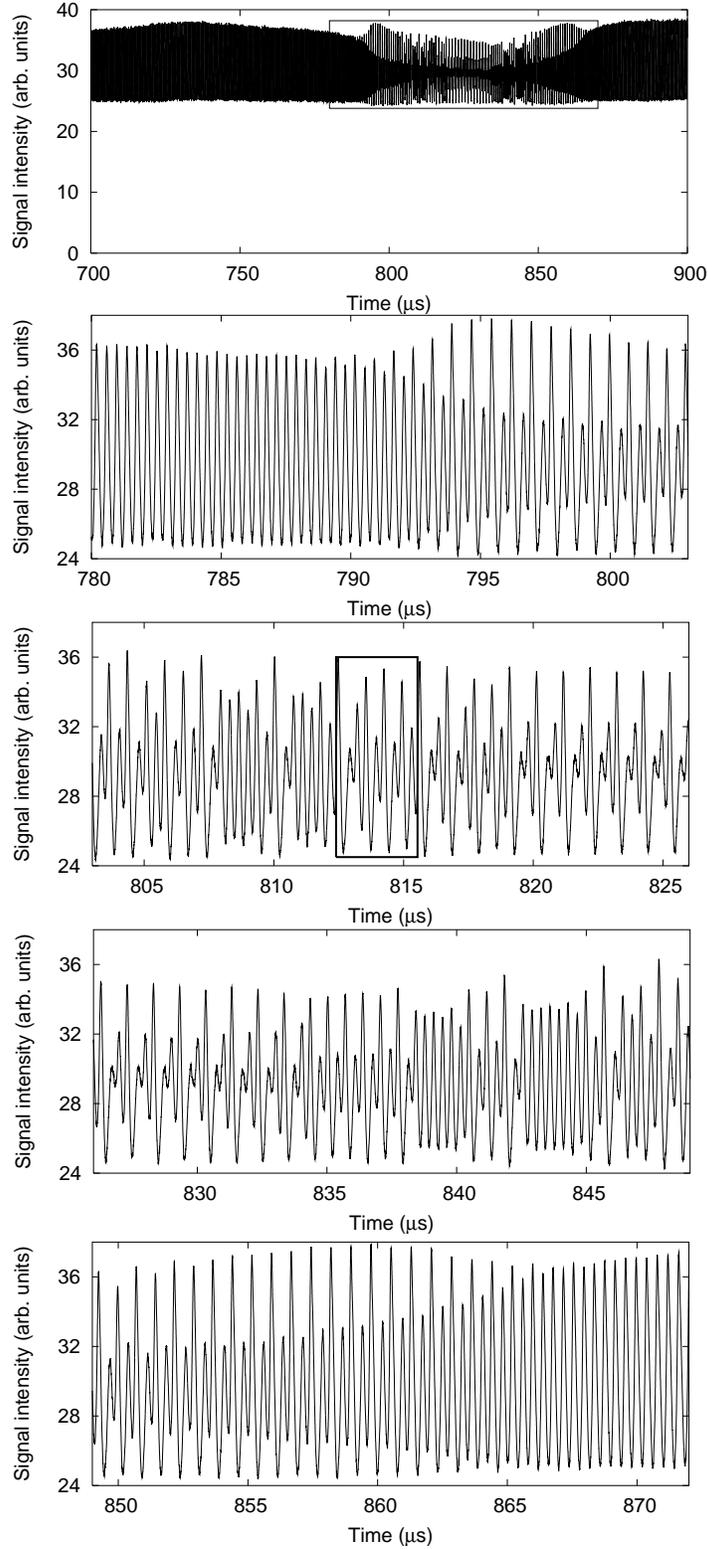}
  \caption{(a) Signal intensity vs. time; (b)-(e)
    consecutive excerpts from the irregular burst occurring for
    $t\in[780,875]$.  The segment between $t=812.421$ and $t=815.562$
    contains a period-9 orbit used in the subsequent topological
    analysis.  }
  \label{fig:sequence}
\end{figure}

The TROPO used in the experiment is as described in
Refs.~\cite{Suret00,Suret01a,Suret01b}. It features a 15-mm-long KTP
crystal cut for type-II phase matching, enclosed inside a 63-mm-long
cavity delimited by two mirrors with a radius of curvature of 50 mm.
Cavity length is not actively stabilized. The cavity is resonant at
532 nm, the wavelength of the frequency-doubled Nd:YAG pump laser, and
at 1064 nm, near which two infrared fields are generated.  Parametric
threshold is reached at pump powers of the order of 10 mW. At a pump
power of 3.5 W (i.e., 350 times above threshold), we have observed in
the output signal intensity waveforms more complex than the periodic
instabilities reported so
far~\cite{richy95,Suret00,Suret01a,Suret01b,amon03:_burst}. The time
series that we analyze in this paper is shown in
Fig.~\ref{fig:sequence}. Because the raw recording had limited
vertical resolution, it has been processed through an acausal lowpass
filter with a cutoff frequency of 7 MHz. Acausal filtering of time
series has been shown not to introduce artifacts in subsequent
phase-space reconstructions~\cite{mitschke90:_acaus}. Between $t=$790
$\mu$s and $t=$870 $\mu$s, Fig.~\ref{fig:sequence}(a) displays a burst
of irregular behavior inside a long interval of quasi-sinusoidal
periodic behavior of frequency approximately 3 MHz. The rapid
variation of the periodic waveform with time shows clearly that the
system is highly nonstationary. The drift occurs on a time scale
consistent with previous reports of thermal effects in this system
\cite{Suret00,Suret01a}.

The complex burst shown in detail in Figs.~\ref{fig:sequence}(b)-(e)
is highly suggestive of deterministic chaos. Between $t=$790 $\mu$s
and $t=$805 $\mu$s, it begins by a progressive transition from the
base period-1 orbit to a period-2 orbit. This is a signature of a
period-doubling bifurcation, the first step of the most ubiquitous
route to chaos. The reverse bifurcation can be seen in
Fig.~\ref{fig:sequence}(e). Moreover, the time series displays a
number of sequences of almost periodic behavior, a hallmark of
low-dimensional chaos~\cite{Gilmore02,Gilmore98}. In particular, the segment
between $t=$830 $\mu$s and $t=$840 $\mu$s contains three
periodic bursts of periods 3, 2, and 1.

\begin{figure}[htbp]
  \centering
  \includegraphics[width=5.5in]{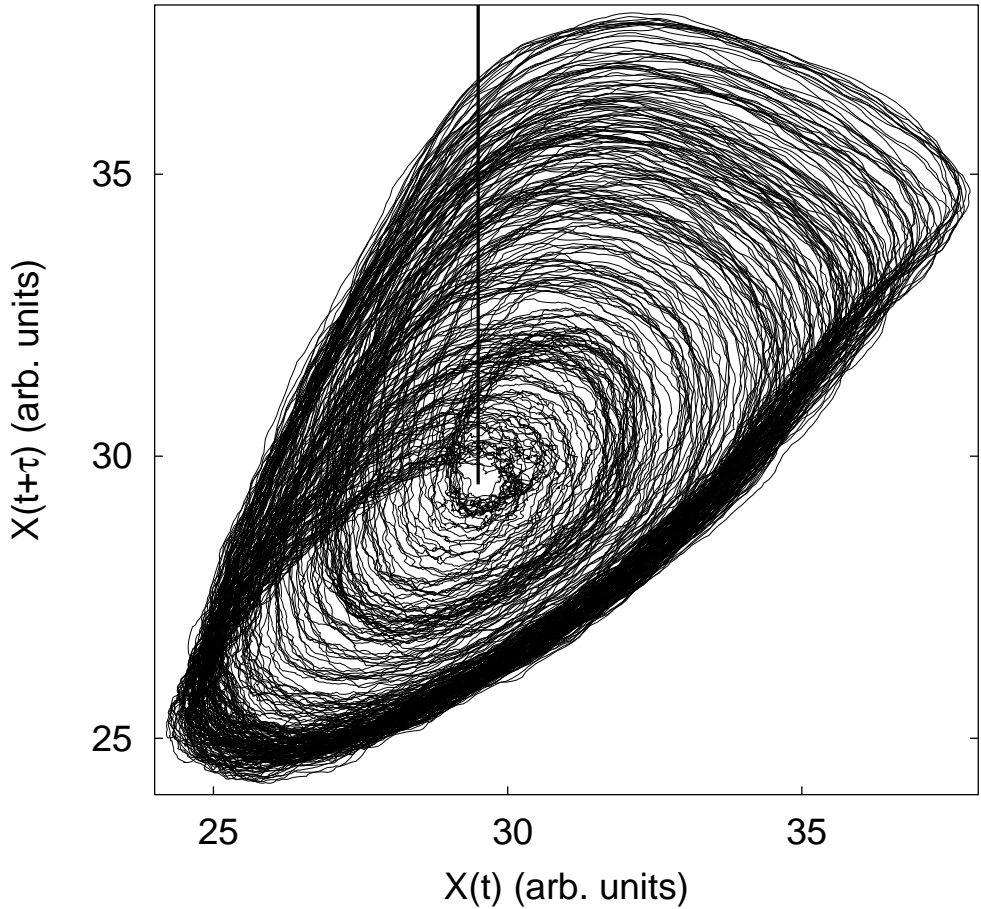}
  \caption{Phase-plane portrait $[X(t),X(t+\tau)]$, with $X(t)$ the
    time series of Figs.~\ref{fig:sequence}(b)-(e) and $\tau=$55 ns.
    The vertical line indicates the section plane used in the
    subsequent analysis. Flow rotates clockwise around the hole in the
    middle.}
  \label{fig:portrait}
\end{figure}

When chaotic behavior is suspected, a natural step is to try to
reconstruct a strange attractor in a phase space using, e.g., the
method of time-delays~\cite{Ott93,Abarbanel96,Gilmore02,Gilmore98}. In the
present case, this procedure is questionable because the system is not
stationary.  Indeed, the superposition of trajectory segments
corresponding to different values of control parameters is expected to
yield a blurred plot. However, a phase-plane plot of the time series
of Fig.\ref{fig:sequence}(b) through \ref{fig:sequence}(e) is surprisingly
similar to a R\"ossler-type chaotic attractor
(Fig.~\ref{fig:portrait}). This indicates that trajectories of our
system change their shape relatively slowly as a control parameter is
varied.

\begin{figure}[htbp]
  \centering
  \includegraphics[width=5.5in]{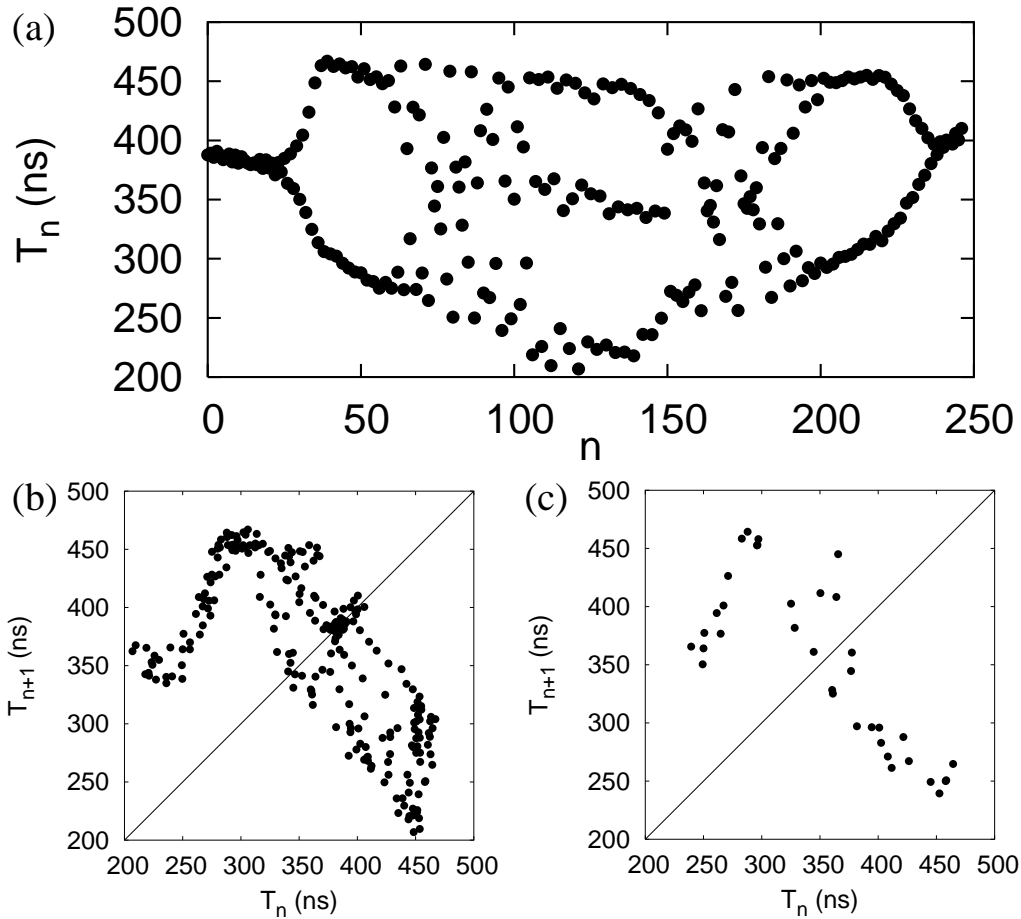}
  \caption{(a) Poincar\'e section return times $T_n$ vs.
  intersection number $n$. Return maps 
    $(T_n,T_{n+1})$ (b) for $n\in[1,250]$; (c) for $n\in[70,105]$.}
  \label{fig:times}
\end{figure}

Next, we choose a Poincar\'e section (Fig.~\ref{fig:portrait}) and
construct a first return map for it. A convenient choice is the return
map for the times of flight $T_n$ between the $n$-th and the
$(n+1)$-th intersections with the section plane, which are relatively
insensitive to noise. How $T_n$ varies along the time series is shown
in Fig.~\ref{fig:times}(a). This plot clearly displays the bifurcation
diagram of a system that undergoes a period-doubling cascade, explores
a chaotic zone followed by a period-3 periodic window and then goes
back. Note that the fraction of the diagram where complex
behavior is observed is relatively small.

The plot $(T_n,T_{n+1})$ is shown in Fig.~\ref{fig:times}(b), where a
folded structure similar to a one-dimensional map can easily be
discerned. This suggests that the irregular dynamics observed in the
experiment is deterministic. However, no quantitative information can
be extracted from Fig.~\ref{fig:times}(b) which is blurred by
variations in control parameters, as can be verified by plotting
separately graphs corresponding to different parts of the time series.
If we restrict ourselves to the first chaotic zone of
Fig.~\ref{fig:times} (i.e., for $n\in [70,105]$, or
$t\in[806.8,819.0]$), the resulting plot is much closer to a
one-dimensional map but there are now too few points to rigorously
assess the presence of deterministic chaos, let alone to quantify it
[Fig.~\ref{fig:times}(c)].

Indeed, most characterization methods require that the neighborhood of
each point in phase space is sufficiently populated to capture the
local structure of the attractor. Thus, they depend on the time series
\emph{nonlocally} in time: nearest neighbors in phase space are
usually located far apart in the time series (\textit{long-term
  recurrence}). This makes these methods fragile with respect to
variations of control parameters along the time series. As we see
below, a topological analysis circumvents this problem by extracting
information from isolated time series segments, namely those
approaching a closed orbit (\textit{short-term recurrence}): we shall
not only extract a signature of an underlying chaotic dynamics in the
fixed-parameter system but also obtain lower bounds on its topological
entropy, a classical measure of chaos.

The time series of Fig.~\ref{fig:sequence} displays many periodic
events. As with the period-3 orbit, many of them correspond to stable
periodic windows explored by the system as parameters are swept but
some can also be found in zones of irregular behavior. They then
likely correspond to a close encounter with one of the infinity of
unstable periodic orbits embedded in a chaotic attractor.

The topological analysis of chaos~\cite{Gilmore02,Gilmore98} proceeds by
characterizing the organization of periodic orbits, which are
associated with closed curves in phase space. In dimension three, how
these curves are intertwined can be studied using knot theory and
branched manifolds (``templates''). This is because the knot
invariants of periodic orbits provide topological signatures of the
stretching and squeezing mechanisms organizing a strange
attractor~\cite{Gilmore02,Gilmore98}. A major advantage of topological analysis
is that each time series segment shadowing a periodic orbit,
\emph{stable or unstable}, can be analyzed independently of the
others.

Although the time series of Figs.~\ref{fig:sequence}(b)-(e) is very
short, we have detected several closed orbits in it. The criterion
was that the orbit should return to its initial condition in the
$(T_n,T_{n+1})$ plane with 3\% accuracy. These orbits are the
period-$1$, period-$2$, period-$4$ and period-$8$ orbits of the
period-doubling cascade, the period-$3$ orbit as well as 4 orbits
of period $6$, $7$, $9$ and $10$.

We have found that the braids associated with these closed orbits can
all be projected to a standard horseshoe template~\cite{Gilmore02,Gilmore98}, as
the structure of the return map suggests. More precisely, they have
the same braid types as orbits $1$, $01$, $01^3$, $01^3(01)^2$,
$01^6$, $011010111$, $(011)^31$, $011$ and $01^5$ of the standard
horseshoe template (ordered as in the time series). Linking numbers
were not computed as this would require comparing different parts of
the time series. This observation suggests that although orbits change
their shape as control parameters vary along the time series, their
topological organization is not modified. Such a robustness would be
extremely unlikely if the irregular dynamics observed was not
deterministic.

\begin{figure}[htbp]
  \centering
  \includegraphics[width=5in]{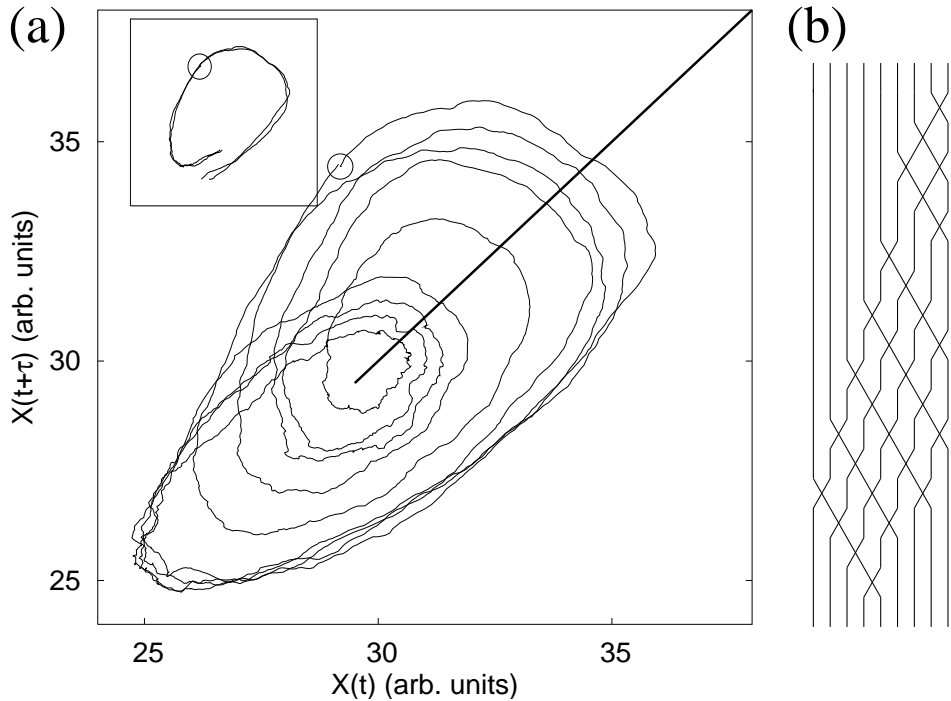}  
  \caption{(a) A period-$9$ closed orbit detected in the time series.
    The open circle indicates its starting and final points. The inset
    displays trajectory segments around these two points; (b) a
    presentation of this orbit as an open braid, using the diagonal as
    a Poincar\'e section. }
  \label{fig:period9}
\end{figure}

However, topological analysis can provide us with stronger evidence.
Indeed, we have found that two of the closed orbits have a
``positive-entropy'' braid type. One is shown in
Fig.~\ref{fig:period9}. In a stationary system, how trajectories are
stretched and folded around such an orbit forces the existence of an
infinity of periodic orbits~\cite{Boyland94:_topol}. Thus,
positive-entropy orbits exist only in systems that have experienced
infinitely many bifurcations and are chaotic in some region of
parameter space~\cite{Boyland94:_topol,Hall94a,Mindlin93a,Gilmore02,Gilmore98}:
a ``pretzel knot'' (i.e., a common type of positive-entropy orbit
having such a knot type) in a three-dimensional flow implies chaos
just as a period-three orbit in a one-dimensional map does.  The
presence of such an orbit in an experimental data set was first used
as an indicator of chaos for the Belousov-Zhabotinskii
reaction~\cite{Mindlin91a}. Pretzel knots have also been observed in
laser experiments~\cite{Papoff92a,Lefranc93a}. How many orbits are
forced by a given braid type is quantified by its topological entropy,
which is a lower bound on the topological entropy of systems
containing it~\cite{Boyland94:_topol}.

Using the Trains program by T. Hall~\cite{Hall:_train}, we have
computed the topological entropy of the braid in
Fig.~\ref{fig:period9}(b) to be $h_T\sim 0.377057 > 0$. Similarly, the
period-10 orbit has entropy $h_T\sim 0.473404 > 0$. If our system were
stationary (as in
Refs.~\cite{Mindlin91a,Papoff92a,Lefranc93a}), the presence of such
orbits would unambiguously imply chaos and provide a lower bound on
topological entropy. However, the closed orbit in
Fig.~\ref{fig:period9} is not a true periodic orbit, since control
parameters have different values at its starting and final points.

Nevertheless, the following observation gives us confidence that the
observed braid type is actually present in the unperturbed system. The
inset of Fig.~\ref{fig:period9} shows two trajectory segments
$\{X(t);t\in[t_0-\tau_1,t_0+\tau_2]\}$ and
$\{X(t);t\in[t_0+T-\tau_1,t_0+T+\tau_2]\}$ centered around the
starting point $X(t_0)$ and the ending point $X(t_0+T)$ of the closed
orbit, with $\tau_1+\tau_2 = 350$ ns being approximately one orbital
period. Although the two segments correspond to the extreme values of
the control parameter along the closed orbit (Fig.~\ref{fig:times}(a)
indicates that parameter variation is monotonic for $n\in[86,93]$), we
see that they are almost superimposed on each other. This indicates
that the vector flow changes very little between $t=t_0$ and
$t=t_0+T$. The separation between the two segments can then be taken
as an upper bound on the separation between the closed orbit observed
and a periodic orbit of the unperturbed system. As this separation is
significantly smaller than the separation between strands of the
closed orbit in Fig.~\ref{fig:period9}(a), the unperturbed system
(i.e., the TROPO with fixed cavity length) must have a periodic orbit
with the braid type shown in Fig.~\ref{fig:period9}(b). We can then
conclude that it exhibits deterministic chaos. Similarly,
perturbations due to noise have no influence if their amplitude is
small with respect to interstrand distance.

Our study shows that closed orbits with a positive-entropy braid type
can be exploited when the influence of parameter variation is small
over one orbit period, a modest requirement compared to other methods.
However, we expect this approach to be also applicable to stronger
nonstationarity. Generically, closed orbits of a system with a swept
parameter connect continuously to periodic orbits of the unperturbed
system as the sweeping rate goes to zero, as is easily shown using the
Implicit Function Theorem. If a positive-entropy closed orbit can be
shown not to change its braid type along the homotopy path, then it
provides a signature of chaos. Understanding when braid type is
preserved is difficult, but a preliminary study of the logistic map
with a swept parameter (where permutation of the periodic points is
the counterpart of braid type) suggests that this approach is quite
robust. Regarding applicability to higher-dimensional systems, where
knots and braids fall apart, we are currently developing alternate
methods for computing the topological entropy of a periodic orbit.

In conclusion, sophisticated topological methods have allowed us to
obtain the first experimental signature of deterministic chaos in an
optical parametric oscillator using only a very short segment of a
nonstationary time series, which contained a positive-entropy orbit.
Moreover, the fact that two such orbits were detected in less than 40
cycles indicates that these signatures of chaos are extremely robust
with respect to variation in control parameters.

We thank R. Gilmore, T. Hall, S. Bielawski, D. Derozier and J.
Zemmouri for stimulating discussions.  CERLA is supported by the
Minist\`ere charg\'e de la Recherche, R\'egion Nord-Pas de Calais and
FEDER.

\end{document}